\documentclass[10pt]{article}
\usepackage[totalwidth=450pt, totalheight=590pt, centering]{geometry}

\usepackage{amsmath,amssymb,mathrsfs}
\usepackage{bbm} 

\usepackage[dvips]{graphicx}

\usepackage{pstricks}
\usepackage{pst-grad}

\usepackage[dvips,bookmarks=false]{hyperref} 
\hypersetup{pdfstartview=FitH,pdfhighlight=/O,colorlinks=false}

\newcommand{\be}{\begin{equation}}
\newcommand{\ee}{\end{equation}}
\newcommand{\ben}{\begin{equation*}}
\newcommand{\een}{\end{equation*}}

\newcommand{\mc}[1]{\mathcal{#1}}

\newcommand{\mbb}[1]{\mathbb{#1}}

\newcommand{\eps}{\varepsilon}

\newcommand{\mdots}{,.\,.\,,}

\title{\LARGE{\textbf{Loop Quantum Gravity \`a la Aharonov-Bohm}}}

\author{Eugenio Bianchi\footnote{email: {\tt bianchi@cpt.univ-mrs.fr}}\\[.35em]
\normalsize{\emph{Centre de Physique Th\'eorique de Luminy{$\,$}\footnote{\scriptsize Unit\'e mixte de recherche (UMR 6207) du CNRS et des Universit\'es de Provence (Aix-Marseille I), de la M\'editerran\'ee (Aix-Marseille II) et du Sud (Toulon-Var); laboratoire affili\'e \`a la FRUMAM (FR 2291).} , case 907, F-13288 Marseille, France}}
}

\date{September 10, 2009}

\begin{document}

\maketitle


\begin{abstract}
The state space of Loop Quantum Gravity admits a decomposition into orthogonal subspaces associated to diffeomorphism equivalence classes of spin-network graphs. In this paper I investigate the possibility of obtaining this state space from the quantization of a topological field theory with \emph{many} degrees of freedom. The starting point is a 3-manifold with a network of defect-lines. A locally-flat connection on this manifold can have non-trivial holonomy around non-contractible loops. This is in fact the mathematical origin of the Aharonov-Bohm effect. I quantize this theory using standard field theoretical methods. The functional integral defining the scalar product is shown to reduce to a finite dimensional integral over moduli space. A non-trivial measure given by the Faddeev-Popov determinant is derived. I argue that the scalar product obtained coincides with the one used in Loop Quantum Gravity. I provide an explicit derivation in the case of a single defect-line, corresponding to a single loop in Loop Quantum Gravity. Moreover, I discuss the relation with spin-networks as used in the context of spin foam models.

\end{abstract}




\section{Introduction}
Background independence and diffeomorphism invariance are key assumptions in various approaches to quantum gravity \cite{Oriti:book2009}. Compared to older canonical approaches \cite{Isham:1992ms}, what makes Loop Quantum Gravity \cite{Rovelli:2004tv,Ashtekar:2004eh,Thiemann:2007zz} technically possible and mathematically robust is a further assumption consisting in the choice of the loop algebra as the basis for quantization\footnote{For an enlightening discussion on the question ``Why loops?'' we refer to the review \cite{Rovelli:2008zza}} \cite{Rovelli:1987df}\cite{Ashtekar:1993wf}. The kinematical Hilbert space of the theory turns out to admit a decomposition into orthogonal subspaces $\mc{K}_\Gamma$ associated to diffeomorphism equivalence classes of graphs $\Gamma$. Spin network states with graph $\Gamma$ \cite{Rovelli:1995ac} provide a complete orthonormal basis of $\mc{K}_\Gamma$. 

The motivation for this paper comes from the following remark: the mathematical structure of the Hilbert space $\mc{K}_\Gamma$ and its physical interpretation in terms of quantum geometries \cite{Rovelli:1994ge,Bianchi:2008es} are the ones proper to the Hilbert space of a system with a finite (but possibly large) number of degrees of freedom. Therefore, while the full theory has an infinite number of degrees of freedom (as classical General Relativity does), actually each subspace $\mc{K}_\Gamma$ captures only a finite number of them. The question we want to explore in this paper is if the state space $\mc{K}_\Gamma$, together with its physical interpretation, can be obtained directly from the quantization of a classical system with a finite number of degrees of freedom. A classical system that is generally considered for this purpose is a discrete gauge theory on an abstract graph $\Gamma$ \cite{Immirzi:1996dr,Barbieri:1997ks,Markopoulou:1997ri,Baez:1999sr}. This model captures some important features of Loop Quantum Gravity. However, as the graph $\Gamma$ is not embedded into a $3$-manifold, in this model the role played by diffeomorphisms is rather obscure. Moreover, potential problems with locality have been identified \cite{Markopoulou:2007ha}. On the other hand, here we look for a model such that a manifold is present and diffeomorphisms act on local fields in the standard way, via pullback.

The classical system  we investigate is a field theory: we consider a theory of locally-flat connections on a manifold. The manifold is assumed to be non simply-connected because of the presence of a network of defects. We quantize this system relying only on standard field theoretical methods \cite{'tHooft:2005cq}. In particular, the quantization procedure we discuss does not involve the basic assumption about loop observables which characterizes Loop Quantum Gravity. Nevertheless, the resulting state space is surprisingly close to the one of ordinary Loop Quantum Gravity. Spin networks arise again and provide the appropriate tool for describing gauge- and diffeomorphism-invariant functionals of the connection. The role played by defects and loops in this approach is analogous to the one played by solenoids and Wilson loops in the Aharonov-Bohm effect \cite{Aharonov:1959fk}. This justifies the title of the paper.


The paper is organized as follows: in section \ref{sec:foundations} we introduce our model, in section \ref{sec:line} we analyze in detail a simple example and in section \ref{sec:perspectives} we discuss some perspectives.

\section{Defects, locally-flat connections and the state space}\label{sec:foundations}

Let $\Sigma$ be a connected $3$-manifold without boundary and consider a network\footnote{By a network we mean a set of points $\{v_r\}\in \Sigma$ and of curves $\{e_i\}$ connecting them. We require that at each of the points $v_r$ at least two curves meet.} $\mc{D}$ of curves embedded in $\Sigma$. In the following we will focus on a particular case: the case where the network $\mc{D}$ arises from the $1$-skeleton of a cellular decomposition $\mc{C}(\Sigma)$ of the manifold $\Sigma$ \cite{Fritsch:1990ct},
\begin{equation}
\mc{D}\equiv \mc{C}_1(\Sigma)\;.
\label{eq:}
\end{equation}
We will regard the set $\mc{D}$ as a network of \emph{defects} in the manifold $\Sigma$. In particular we introduce a new manifold $\Sigma'$ obtained by subtracting the network $\mc{D}$ from the original manifold $\Sigma$,
\begin{equation}
\Sigma'=\Sigma-\mc{D}\;.
\end{equation}
The manifold $\Sigma'$ is path-connected but not simply-connected. Closed paths around edges of the $1$-skeleton $C_1(\Sigma)$ are non-contractible, therefore the first homotopy group $\pi_1(\Sigma')$ is non-trivial\footnote{For notational conventions and a discussion of a presentation of the group $\pi_1(\Sigma-\mc{D})$ we refer to appendix \ref{app:pi1}.}.


On the manifold $\Sigma'$ we can consider a locally-flat connection $A(x)$ for some gauge group $G$. Despite being locally-flat, the connection can have non-trivial holonomy around non-contractible loops in $\Sigma'$. This is in fact the mathematical origin of the Aharonov-Bohm effect. We call $\mc{A}_f$ the space of locally-flat connections
\begin{equation}
\mc{A}_f=\{A(x) \;|\; F(A)=0\;\; \text{on}\;\; \Sigma'\}
\end{equation} 
and $\mc{A}_f/\mc{G}$ the space of locally-flat connections modulo gauge transformations. The space $\mc{A}_f/\mc{G}$ is finite dimensional and its elements are completely characterized by their holonomy around non-contractible loops. We have in fact that a configuration in $\mc{A}_f/\mc{G}$ can be specified in terms of a homomorphism from $\pi_1(\Sigma')$ into $G$, up to conjugation by $G$. We call $\mc{N}$ this \emph{moduli} space,
\begin{equation}
\mc{N}=\text{Hom}(\pi_1(\Sigma'),G)/G\;,
\end{equation}
and $\{m_r\}$ coordinates (moduli) on $\mc{N}$. More explicitly, a locally-flat connection is labeled by its moduli $m_r$ and a gauge transformation $g:\Sigma'\rightarrow G$,
\begin{equation}
A^{m_r,g}=g^{-1}\bar{A}^{m_r}g+g^{-1}dg\;,
\label{eq:Amg}
\end{equation}
with the connection $\bar{A}^{m_r}$ satisfying a gauge fixing condition $\chi(\bar{A}^{m_r})=0$.

Now we consider the kinematics of General Relativity in Ashtekar-Barbero variables  \cite{Ashtekar:1986yd}. The configuration variable is a $SU(2)$ connection $A(x)$ on a $3$-manifold $\Sigma$. A basic step in attempts to canonically quantize General Relativity involves the introduction of a kinematical Hilbert space $\mc{K}$ of functionals of the connection. Such functionals are required to be invariant under $SU(2)$-gauge transformations and under diffeomorphisms $\text{Diff}(\Sigma)$ of the manifold $\Sigma$. Loop Quantum Gravity provides a complete implementation of this step. Here we follow a similar procedure with two specific differences:
\begin{itemize}
	\item we look for a Hilbert space $\mc{K}'$ of states $\Psi[A]$ associated to the 3-manifold $\Sigma'$, instead of $\Sigma$;
	\item besides $SU(2)$-gauge invariance and $\text{Diff}(\Sigma')$ invariance, we require \emph{topological} invariance of the states $\Psi[A]\,$. The condition can be imposed \`a la Dirac as a constraint,
\begin{equation}
\hat{F}\, \Psi[A] =0\;.
\label{eq:top constraint}
\end{equation}
Here $\hat{F}$ is the operator associated to the curvature of the connection.
\end{itemize}  
As well known, in a topological gauge theory \cite{Witten:1988ze,Witten:1988hf,Horowitz:1989ng}\cite{Birmingham:1991ty,Cattaneo:1995tw} gauge invariance and diffeomorphism invariance are strictly related. Therefore we can focus on gauge invariant functionals which satisfy the topological-invariance constraint (\ref{eq:top constraint}) only. Diffeomorphism invariance comes for free. Equivalently, 
 we can consider the space of gauge invariant functionals of a locally-flat connection.

States belonging to the state space $\mc{K}'$ described above depend on the connection $A(x)$ only through its moduli $\{m_r\}\in \mc{N}$. Therefore we have that our functionals are labeled by a function $f:\mc{N}\rightarrow \mbb{C}$ so that
\begin{equation}
\Psi_f[A^{m_r,g}]=f(m_1\mdots m_R)\;.
\end{equation}  
There is another characterization of gauge invariant functionals of a flat connection which is slightly redundant, but definitely clearer. Let $\Gamma$ be a graph embedded in $\Sigma'$. We call $\{\gamma_1\mdots \gamma_L\}$ its links and introduce the holonomy along a link of $\Gamma$ as the functional $h_{\gamma}:\mc{A}_f\rightarrow SU(2)$ defined in the standard way,
\begin{equation}
h_\gamma[A]=\text{P}\exp \int_\gamma A\;.
\end{equation}
Then we introduce a class of gauge invariant functionals of a locally-flat connection: these functionals are labeled by a graph $\Gamma$ and by a function $\eta:SU(2)^L\rightarrow \mbb{C}$. The functionals $\Psi_{\Gamma,\eta}[A]$ are given by
\begin{equation}
\Psi_{\Gamma,\eta}[A]=\eta(h_{\gamma_1}[A]\mdots h_{\gamma_L}[A])\;
\label{eq:cylGamma}
\end{equation}
and the function $\eta$ is assumed to be invariant under $SU(2)$ conjugation at nodes, $g_s h_\gamma g^{-1}_{t}$, in order to guarantee the gauge invariance of the functional $\Psi_{\Gamma,\eta}[A]$. A remark is in order at this point: as the connection we are considering is locally flat, we have that the functional $\Psi_{\Gamma,\eta}[A]$ depends on the graph $\Gamma$ only via its `knotting' with the skeleton $C_1(\Sigma)$. In particular, such states are invariant under diffeomorphisms $\phi\in\text{Diff}(\Sigma')$ connected to the identity,
\begin{equation}
\Psi_{\Gamma,\eta}[\phi^*A]=\Psi_{\phi^{-1}\circ\Gamma,\eta}[A]=\Psi_{\Gamma,\eta}[A]\;.
\end{equation}
As a result we have that different couples $(\Gamma,\eta)$ can actually represent the same state. For instance, for a closed curve $\gamma$ which is contractible in $\Sigma'$, the Wilson loop state $\Psi_{\gamma, \text{Tr}_j}[A]=\text{Tr} D^{(j)}(h_\gamma[A])$ evaluates to the identity  and represents the same state as $\Psi_0[A]=1$. In this sense the couples $(\gamma,\text{Tr}_j)$ and $(\emptyset,1)$ correspond to the same state. To check if two functionals labeled by different couples $(\Gamma,\eta)$ represent in fact the same state, we can always go back to the non-redundant description in terms of moduli. In particular we have that there is a unique function $f$ on $\mc{N}$ such that 
\begin{equation}
\Psi_{\Gamma,\eta}[A^{m_r,g}]=f(m_1\mdots m_R)
\end{equation}
for the whole class of equivalent states having different couples $(\Gamma,\eta)$. 

In the specific case of a defect-network $\mc{D}$ arising from the $1$-skeleton of a cellular decomposition of $\Sigma$, there is a class of graphs $\Gamma$ of particular interest that we describe here: it is the class of graphs $\Gamma'$ \emph{dual} to the cellular decomposition $\mc{C}(\Sigma)$,
\begin{equation}
\Gamma'= {\mc{C}(\Sigma)^*}_1\;.
\end{equation}
By dual we mean that in each cell of the decomposition $\mc{C}(\Sigma)$ we have a node of the graph and that two nodes are connected by a path (a link of the graph) if the two corresponding cells are adjacent\footnote{The link is assumed to be a subset of the region given by the union of the two cells.}. We notice that \emph{loops} of the graph $\Gamma$ encircle once the edges of the $1$-skeleton of $C(\Sigma)$ so that the edge-path group $\pi(\Gamma')$ of the graph $\Gamma'$ is isomorphic to the first homotopy group of the manifold $\Sigma'$,
\begin{equation}
\pi(\Gamma')\sim \pi_1(\Sigma')\;.
\end{equation}
As a result we can describe the moduli space $\mc{N}$ in terms of homomorphisms from the edge-path group of the graph to SU(2), modulo conjugation,
\begin{equation}
\mc{N}=\text{Hom}(\pi(\Gamma'),G)/G\;.
\end{equation}
States of the form (\ref{eq:cylGamma}), but with graph $\Gamma'$, now actually depend only on the function $\eta$. The relation between the function $\eta$ and the function $f$ can be derived via a gauge-fixing of the $SU(2)$ invariance at the nodes of the graph, for instance setting to the identity the group elements $h_\gamma$ associated to links $\gamma$ belonging to a maximal tree on $\Gamma'$.\\

In order to promote the linear space described above to a Hilbert space we need to introduce a scalar product. A topologically invariant functional measure $\mc{D}[A]$ can be introduced via the requirement that it reduces to an ordinary measure on $\mc{N}$,
\begin{equation}
\langle f | g\rangle=\int_{\mc{A}_f/\mc{G}}\!\! \mc{D}[A]\;  \overline{\Psi_f[A]}\;\Psi_g[A]=\int_{\mc{N}}  d\mu(m_r)\;\overline{f(m_1\mdots m_R)}\;g(m_1\mdots m_R)\;.
\label{eq:measure0}
\end{equation}  
The Hilbert space of states $\mc{K}'$ associated to $\Sigma'$ reduces to a Hilbert space $L^2(\mc{N},d\mu)$ of square integrable functions on the moduli space. Making sense of the formal functional measure $\mc{D}[A]$ on the left hand side of (\ref{eq:measure0}) amounts to a choice of measure $d\mu(m_r)$ on the moduli space $\mc{N}$. As a result, the following problem arises: is there a principle that can guide us in the choice of the measure $d\mu(m_r)$?

Our aim here is to use the methods proper to gauge field theory \cite{'tHooft:2005cq} to make sense of the functional integral over locally-flat connections modulo gauge transformations. Such methods, in the case of a finite dimensional moduli space, allow to fully determine a specific measure on the moduli space\footnote{These techniques are the ones used for the string measure \cite{Polyakov:1981rd} and for the path integral quantization \cite{Carlip:1995jn} of $2+1$ dimensional gravity with and without point particles \cite{Witten:1988hc}. We point out also a close relationship with works on the field theoretical measure for simplicial geometries  \cite{Jevicki:1985ta}.}  and therefore answer our question. The guiding principle here is \emph{locality}, that is: the configurations we are integrating over are local fields. The integral over locally-flat connections $\mc{D}[A^{m_r,g}(x)]$ can be thought of as an integral over the moduli $m_r$ \emph{and} over gauge transformations $g(x)$. As we are considering only integrals of gauge invariant quantities, the integral over gauge transformations can be factorized and we end up with a finite dimensional integral over moduli, with a non-trivial measure on the moduli. Such measure comes from the Jacobian of the change of variables on configuration space. This geometrical approach is equivalent \cite{Bern:1990bh} to the more standard and well-known Faddeev-Popov procedure that we follow here. 

The measure on locally-flat connections modulo gauge transformations can be formally written as follows
\begin{equation}
\int_{\mc{A}_f/\mc{G}}\mc{D}[A]\;=\int_{\mc{A}} \Big[\prod_x dA(x)\Big]\,\delta\big(F(A)\big)\,\delta\big(\chi(A)\big)\;\Delta_{\text{FP}}(A)\;.
\end{equation}
Here $\mc{A}$ is the space of connections over $\Sigma'$, $\delta\big(F(A)\big)$ imposes that the connection is locally flat, $\delta\big(\chi(A)\big)$ imposes a gauge fixing condition $\chi(A)=0$ on the connection, and $\Delta_{\text{FP}}(A)$ is the standard Faddeev-Popov determinant
\begin{equation}
\Delta_{\text{FP}}(A)=\Big|\text{Det} \frac{\delta \chi}{\delta \xi}\Big|\;,
\label{eq:FP def}
\end{equation}
i.e. the determinant of the variation of the gauge fixing condition under a small gauge transformation with parameter $\xi$. As well-know, such expression is (formally) independent from the choice of gauge fixing condition $\chi$. Now, the delta function with support on flat connections can be integrated out going to the locally-flat connections $A^{m_r,g}(g)$ of expression (\ref{eq:Amg}). Similarly, the delta function with support on connections satisfying the gauge choice $\chi=0$ can be integrated out going to the gauge-fixed locally-flat connections $\bar{A}^{m_r}$ which depend only on the moduli $m_r$. As a result we end up with the following finite dimensional integral with a non-trivial measure
\begin{equation}
\int_{\mc{A}_f/\mc{G}}\mc{D}[A]\;=\int d\bar{A}^{m_r} \;\Delta_{\text{FP}}(\bar{A}^{m_r})\;
=\int_{\mc{N}}dm_1\cdots dm_R\;J(m_r) \Delta_{\text{FP}}(m_r)\;.
\end{equation}
The function $J(m_r)$ is the Jacobian from gauge-fixed connections $\bar{A}^{m_r}$ to the moduli $m_r$. A procedure for computing it is described in appendix \ref{app:J}. By the function $\Delta_{\text{FP}}(m_r)$ we simply mean the Faddeev-Popov determinant evaluated on the connection $\bar{A}^{m_r}$,
\begin{equation}
\Delta_{\text{FP}}(m_r)=\Delta_{\text{FP}}(\bar{A}^{m_r})\;.
\label{eq:FP moduli}
\end{equation}
This is our field theoretical proposal for the measure $d \mu(m_r)$ in expression (\ref{eq:measure0}) and thus for the Hilbert space $\mc{K}'$: states in $\mc{K}'$ are assumed to be normalizable in the following scalar product
\begin{equation}
\langle f | g\rangle=\int_{\mc{A}_f/\mc{G}}\!\! \mc{D}[A]\;  \overline{\Psi_f[A]}\;\Psi_g[A]=\int_{\mc{N}}  dm_1\cdots dm_R\;J(m_r)\, \Delta_{\text{FP}}(m_r)\;\overline{f(m_1\mdots m_R)}\;g(m_1\mdots m_R)\;.
\label{eq:measureFP}
\end{equation}
In section \ref{sec:line} we investigate this proposal in a rather simple case and provide an explicit expression for the measure $d\mu(m_r)$.\\

The field theoretical derivation of the measure $d\mu(m_r)$ described above can be compared to a different construction which is proper to Loop Quantum Gravity. We describe it in the following. Using the parametrization of states in terms of the class of graphs $\Gamma'$ dual to the cellular decomposition, we have that a natural choice of measure is the Haar measure on links of the graph:
\begin{equation}
\langle \eta | \xi\rangle=\int_{\mc{A}_f/\mc{G}}\!\! \mc{D}[A]\;  \overline{\Psi_{\Gamma',\eta}[A]}\;\Psi_{\Gamma',\xi}[A]=\int_{SU(2)^L}  \prod_{l=1}^L d\mu_H(h_l)\;\overline{\eta(h_1\mdots h_L)}\;\xi(h_1\mdots h_L)\;.
\label{eq:measure haar}
\end{equation}
With this choice of measure, thanks to the Peter-Weyl theorem, we know that \emph{spin network} states with graph $\Gamma'$ provide an orthonormal basis of the Hilbert space $\mc{K}'$. For completeness we recall that the spin network basis corresponds to a specific choice of function $\eta(h_1\mdots h_L)$ labeled by a $SU(2)$ representation (a \emph{spin}) for each link, and an invariant map on the tensor product of representations (an \emph{intertwiner}) for each node. More explicitly\footnote{The notation $\cdot$ stands for a contraction of dual spaces as prescribed by the connectivity of the graph. We recall that, if $V^{(j)}$ is the vector space where the representation $j$ of $SU(2)$ acts, then the tensor product of representations $D^{(j_l)}(h_l)$ lives in $\otimes_l (V^{(j_l)*}\otimes V^{(j_l)})$ while the tensor product of intertwiners lives precisely in the dual of this space.} 
\begin{equation}
\eta_{j_l,i_n}(h_1\mdots h_L)=\Big(\bigotimes_{n\subset\Gamma'} v_{i_n}\Big)\cdot\Big( \bigotimes_{\gamma_l\subset\Gamma'}D^{(j_l)}(h_l)\Big)\;.
	\label{eq:spin network}
\end{equation}
Choosing an orthonormal basis in the intertwiner vector spaces, we have that the orthonormality of spin network states follows,
\begin{equation}
\langle \eta_{j_l,i_n} | \eta_{j'_l,i'_n}\rangle=\big(\prod_l\delta_{j_l,j'_l}\big) \big(\prod_n\delta_{i_n,i'_n}\big)\;.
\end{equation}
It is important to notice that, for functions $\eta$ and $\xi$ invariant under conjugation at nodes, the Haar measure $\prod_l d\mu_H(h_l)$ on $SU(2)^L$ in (\ref{eq:measure haar}) reduces to a specific measure $d\mu(m_r)$ on moduli via the Weyl integration formula \cite{Kirillov:1994rh}. Therefore it provides a definition of the functional measure in (\ref{eq:measure0}), prescribing a specific measure $d\mu(m_r)$ on moduli space.\\

While the group theoretical choice of measure described above is certainly a natural one, it of interest to investigate its relation to the field theoretical choice described before. In the following section we investigate these two proposals in a rather simple case and provide an explicit expression for the measure $d\mu(m_r)$. The result supports the following conjecture:
\begin{itemize}
	\item[] \emph{the measure} (\ref{eq:measureFP}) \emph{on the moduli space $\mc{N}$ obtained via the field theoretical construction coincides with the one obtained from the product of Haar measures} (\ref{eq:measure haar}) \emph{via Weyl integration formula}.
\end{itemize}
If such conjecture turns out to be robust, then we have that spin network states with embedded graph $\Gamma'$ provide an orthonormal basis of the Hilbert space $\mc{K}'$ built via the standard field theoretical construction. 

\section{An example: the state space for a single line defect}\label{sec:line}
In this section we consider the simplest possible example: we assume that the topology of the original manifold $\Sigma$ is trivial, $\Sigma\approx\mbb{R}^3$, and that the defect-network $\mc{D}$ consists of a single line $l\approx\mbb{R}$. The line $l$ is assumed to be unknotted in $\Sigma$. Then, following the construction described in the previous section, we introduce a new manifold $\Sigma'$ obtained from $\Sigma$ subtracting the line $l$:
\begin{equation}
\Sigma'=\Sigma-l\;.
\end{equation}
The manifold $\Sigma'$ has a non-trivial fundamental group $\pi_1(\Sigma')$ generated by the homotopy class $[\gamma]$ of loops  which encircle once the line $l$.

A flat connection in $\Sigma'$ is locally pure gauge, $A_a(x)=g^{-1}(x)\partial_a g(x)$. Here we are interested in identifying the moduli space of such connections modulo gauge transformations. In order to identify a representative in each gauge orbit, we proceed introducing a gauge fixing condition. Such condition has to break the local symmetries of the problem: a convenient choice is to introduce an auxiliary metric $q_{ab}(x)$ on $\Sigma$ and consider the Coulomb-like gauge $\chi=q^{ab}\partial_a A_b$. In order to simplify the calculation we can assume that the line $l$ is a geodesic in this metric. We assume also that the metric is Euclidean so that, choosing Cartesian coordinates\footnote{In Cartesian coordinates, the auxiliary metric is simply $\delta_{ab}$ so that in the following we will raise and lower space indices freely.} $x^a=(x,y,z)$, the line is so to say straight and coincides with the $z$-axes. A locally-flat connection in $\Sigma'$, satisfying the gauge fixing condition
\begin{equation}
\chi^i=\partial^a A_a^i=0\;,
\label{eq:chi}
\end{equation}
is given by
\begin{equation}
A_a^i(x)=\frac{\Phi^i}{2\pi}\alpha_a(x)
\end{equation}
with $\alpha_a(x)$ given by
\begin{equation}
\alpha_a(x)=\left(\frac{-y}{x^2+y^2},\frac{x}{x^2+y^2},0\right)\;.
\end{equation}
This is in fact the general solution\footnote{Up to residual gauge transformations satisfying $\partial^a\partial_a \xi^i=0$.}. We have that its holonomy along a loop $\gamma$ is simply given by
\begin{equation}
h_\gamma[A]=\exp[ i \Big(\int_\gamma \alpha_a dx^a\Big) \frac{\Phi^i}{2\pi} \tau_i\,]\;.
\end{equation}
The integral $\int_\gamma \alpha_a dx^a$ is $2\pi$ times the winding number of the loop $\gamma$ around the line $l$. As a result the holonomy $h_\gamma[A]$ provides a homomorphism from $\pi_1(\Sigma')$ to $SU(2)$. 

Now we would like to provide a physical interpretation for the parameters $\Phi^i$. We introduce the non-abelian magnetic field $B^a_i=\frac{1}{2} \eps^{abc}F_{bc}^i$. As the connection is locally flat in $\Sigma'$, we have that the magnetic field vanishes everywhere in $\Sigma'$ and has support only on the line $l$. It is given by
\begin{equation}
B^a_i(x)=\int_l ds\; \Phi_i \;\dot{x}^a(s)\,\delta^{(3)}(x-x(s))\;.
\end{equation}
\begin{figure}%
\begin{center}
\includegraphics[width=0.4\textwidth]{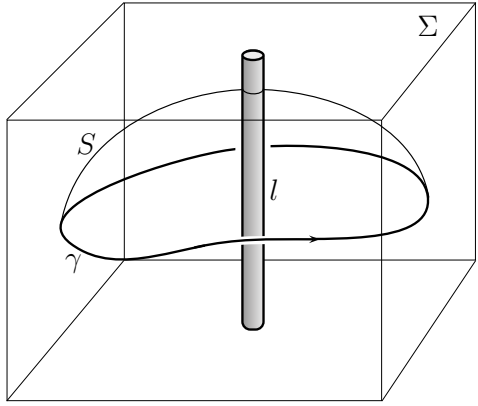}
\end{center}
\caption{A portion of the manifold $\Sigma'=\Sigma-l$. The defect-network consists of the single line $l$. As the connection is locally-flat, the magnetic field vanishes everywhere except along the line $l$. The line can be thought as a thin solenoid which confines the magnetic field.  The flux of the magnetic field through a surface $S$ (in figure) determines the holonomy of the connection along a curve $\gamma$ bounding the surface.}%
\label{pic:solenoid}%
\end{figure}

\noindent We can compute the flux of the magnetic field through a surface $S$ which intersects the line $l$ as in figure \ref{pic:solenoid}. We have that
\begin{align}
\mc{F}_i[B,S]=&\;\int_S\;B^a_i(X(\sigma))\,\eps_{abc}\frac{\partial X^b}{\partial \sigma^1}\frac{\partial X^c}{\partial \sigma^2}\;d\sigma^1 d\sigma^2\\
=&\;\int_l ds\int_S d\sigma^1 d\sigma^2\;\Phi_i\; \eps_{abc}\dot{x}^a(s)\frac{\partial X^b}{\partial \sigma^1}\frac{\partial X^c}{\partial \sigma^2} \delta^{(3)}(X(\sigma)-x(s))\; =\;\Phi_i
\end{align}
Therefore the parameters $\Phi_i$ describe in fact the flux of the magnetic field through the defect line $l$, which can be thought of as a thin \emph{solenoid}. Despite the fact that the magnetic field vanishes outside the solenoid, there are non-local observable effects in the region $\Sigma'$. Such effects are captured by the Wilson loop, i.e. by the holonomy of the connection around non-contractible closed paths. The non-abelian Stokes theorem assures in fact that the exponential of the flux through a surface $\exp i \mc{F}_i[B,S] \tau^i$ coincides with the holonomy of the connection along the loop given by the boundary of the surface, $\gamma=\partial S$.  

The magnetic flux $\Phi^i$ provides a parametrization of the moduli space of locally-flat connections. To find its range we have to identify different magnetic fields which correspond to the same holonomy. Thus the parameter space has the topology of the unit three-sphere $S^3$, i.e. of the group manifold of $SU(2)$.
Therefore, functionals of a locally-flat connection in $\Sigma'$ actually depend only on the flux $\Phi^i$. Moreover, from the requirement that such functionals are gauge invariant, we have that they can depend only on the modulus $\phi$ of the non-abelian flux $\Phi^i$. In fact a global $SU(2)$ transformation rotates the flux, $g^{-1} \Phi^i\tau_i g= {R^i}_j \Phi^j \tau_i$. As a result, states in $\mc{K}'$ are labeled by a function $f(\phi)$
\begin{equation}
\Psi_f[A^{\phi,g}]=f(\phi)\;,
\end{equation}
and the moduli space is given by
\begin{equation}
\mc{N}=\{\Phi^i\in S^3\}/SU(2)=\{\phi\in [0,2\pi]\}\;.
\end{equation}
Now we would like to investigate the field theoretical proposal (\ref{eq:measureFP}) for the measure. In order to determine the measure $d\mu(\phi)$ on the moduli space, we have to compute the Jacobian $J(\phi)$ and the Faddeev-Popov determinant $\Delta_{\text{FP}}(\phi)$. The Jacobian can be easily computed going from Cartesian to spherical coordinates, $d^3\Phi^i=\phi^2 d\phi d^2 v^i$, so that $J(\phi)=\phi^2$. For a more systematic derivation we refer to appendix \ref{app:J}. The Faddeev-Popov term $\Delta_{\text{FP}}$ is given by the determinant of an operator $K(\Phi^i)$ defined as the functional derivative of the gauge fixing condition (\ref{eq:chi}) with respect to a gauge parameter. We have 
\begin{equation}
K_{ij}(\Phi^i)=-\delta_{ij} \Delta-\eps_{ijk} \frac{\Phi^k}{2\pi} \alpha^a\partial_a\;.
\end{equation}
Its eigenfunctions and eigenvalues can be easily found (see appendix \ref{app:FP}). In particular we are interested in the eigenvalues $\lambda_n(\phi)$. They are given by
\begin{equation}
\lambda_n=n^2+n\frac{\phi}{2\pi}\;,
\end{equation}
with $n=\pm 1,\pm 2, \ldots$ and are twice degenerate. As a result, the $\Phi^i$ dependence of the Faddeev-Popov determinant can be extracted considering the appropriately regularized ratio
\begin{equation}
\Delta_{\text{FP}}(\phi)=c\, \frac{\text{Det} K(\Phi^i)}{\text{Det} K(0)}=c\, \frac{\prod_{n=1}^{\infty} \big(\lambda_n(\phi)\big)^2 \big(\lambda_{-n}(\phi)\big)^2}{\prod_{n=1}^{\infty} \big(\lambda_n(0)\big)^2 \big(\lambda_{-n}(0)\big)^2}=c\, \left(\prod_{n=1}^{\infty}\Big(1-\big(\frac{\phi}{2\pi n}\big)^2\Big)\right)^2=c \,\left(\frac{\sin\phi/2}{\phi/2}\right)^2\;,
\end{equation}
where in the last equality we have used the product representation of the sine. The constant $c$ is undetermined and is fixed in the following in such a way that the measure of $\mc{N}$ is normalized to one. Finally we have that the scalar product on $\mc{K}'$ defined by this measure is
\begin{align}
\langle f| g\rangle=& \int_{\mc{A}_f/\mc{G}}\mc{D}[A]\overline{\Psi_f[A]}\Psi_g[A] 
= \int_{\mc{N}}d\phi\; J(\phi)\Delta_{\text{FP}}(\phi)\;\overline{f(\phi)}\,g(\phi)\\ 
=&\frac{1}{\pi}\int_0^{2\pi}d\phi\; \sin^2\frac{\phi}{2}\;\;\overline{f(\phi)}\,g(\phi)
\end{align}
This measure can be compared to the one obtained from the group theoretical construction (\ref{eq:measure haar}) via Weyl integration formula. In this setting, states are labeled by the homotopy class $[\gamma]$ of loops encircling once the line $l$, and by a complex-valued function $\eta$ on $SU(2)$,
\begin{equation}
\Psi_{\gamma,\eta}[A]=\eta(h_{\gamma}[A])\;.
\end{equation}  
Gauge invariance at the base point of the loop $\gamma$ requires that $\eta$ is a class function, $\eta(g^{-1} h g)= \eta(h)$. The scalar product can then be written in terms of the Haar measure on $SU(2)$ and reduces to an integral over the \emph{class angle} $\phi/2$. Defining $f(\phi)=\eta(\exp i \phi\tau_3)$, we have
\begin{align}
\langle \eta| \xi\rangle=& \int_{\mc{A}_f/\mc{G}}\mc{D}[A]\;\overline{\Psi_{\gamma,\eta}[A]}\Psi_{\gamma,\xi}[A] 
= \int_{SU(2)}d\mu(h_\gamma) \;\overline{\eta(h_\gamma)}\,\xi(h_\gamma)\\ 
=&\frac{1}{\pi}\int_0^{2\pi}d\phi\; \sin^2\frac{\phi}{2}\;\;\overline{f(\phi)}\,g(\phi)
\end{align}
which coincides with the field theoretical one derived above. This provides support to our conjecture and leads to a physical interpretation of the class angle as the modulus of the flux of the magnetic field through the defect line.

\section{Perspectives}\label{sec:perspectives}
The analysis we have presented can be developed in a number of directions. In this section we point out some of them.
\begin{itemize}
	\item \emph{Momenta}. In Ashtekar-Barbero variables the momentum conjugate to the connection $A_a^i(x)$ is a non-abelian electric field $E^a_i(x)$. In the setting described in this paper, the connection is assumed to be locally flat and thus labeled by a finite number of moduli $m_r$. Consistently, the electric field $E^a_i(x)$ is labeled by momenta $p^r$ canonically conjugate to the moduli. Therefore, the reduced phase space of the system is given by the cotangent bundle to the moduli space, $\mc{P}=T^*\mc{N}$. At the quantum level, the canonical couple $(m_r,p^r)$ can be promoted to self-adjoint operators satisfying canonical commutation relations. In particular, we have that in the Schroedinger representation the momenta $p^r$ are represented as differential operators
\begin{equation}
\langle f|p^r|g\rangle=\int_\mc{N}dm_r\,J(m_r)\Delta_{\text{FP}}(m_r)\,\overline{f(m_r)}\, \hat{p}^r g(m_r)\;,
\end{equation}
with $\hat{p}^r$ given by
\begin{equation}
\hat{p}^r=-i\frac{\,\partial}{\partial m_r}-\frac{i}{2}\frac{\,\partial}{\partial m_r}\big(\log J(m_r)\Delta_{\text{FP}}(m_r)\big)\;.
\end{equation}
In the case of a single defect line discussed in section \ref{sec:line}, we have that the momentum conjugate to the modulus $\phi$ is given by
\begin{equation}
\hat{p}=-i\frac{\,\partial}{\partial \phi}-\frac{i}{2}\cot\frac{\phi}{2}\;,
\end{equation}
and that the spectrum of the operator $\hat{p}^2$ is discrete. The eigenvalues are labeled by an half integer $j$ and are given by 
\begin{equation}
a_j=j(j+1)+\frac{1}{4}\;.
\end{equation}
This result is to be related to the spectrum of the area of a surface with boundary on the defect-line, as discussed in the following two points.
	\item \emph{Electric field and surfaces}. A basic ingredient in Loop Quantum Gravity is the flux of the electric field through a surface
\begin{equation}
E_i(S)=\int_S \eps_{abc}E^a_i\, dx^b\wedge dx^c\;.
\end{equation}
Within the setting described in this paper, surfaces $S$ belonging to the $2$-skeleton of the cellular decomposition play a distinguished role. The reason is the following. Let us consider two adjacent cells $\mc{R}_1$ and $\mc{R}_2$ in a cellular decomposition $\mc{C}(\Sigma)$. We call $S$ the surface shared by the two regions and $D$ the portion of the $1$-skeleton given by the boundary of this surface, $D=\partial S$. Then we consider a second surface $S'$ diffeomorphic to $S$ and having the same boundary\footnote{For simplicity, we assume that the two surfaces do not intersect, so that there is a region $B$ with the topology of a ball such that its boundary is the surface $S\cup S'$.} $D$. We have that the fluxes $E_i(S)$ and $E_i(S')$ coincide (up to a global $SU(2)$ transformation). This is a consequence of the Gauss constraint $G_i=D_a E^a_i=0$, together with the fact that the connection is locally-flat and the region we are considering is simply-connected. Therefore, quantities which are gauge- and diffeomorphism-invariant can be build up from the fluxes and the holonomy. An example is the quantity $T(S_1,S_2,\gamma)=D(h_\gamma)^{ij}E_i(S_1) E_j(S_2)$. The quantization of these observables is straightforward once they are written in terms of the moduli and of their conjugate momenta $p^r$.
\item \emph{The dual picture of quantum geometry}. In the Ashtekar-Barbero formulation of General Relativity, the electric field plays the role of spatial metric: we have that $\delta^{ij}E^a_i(x)\,E^b_j(x)=h(x) h^{ab}(x)$, where $h_{ab}(x)$ is a Riemannian metric on a $3$-manifold $\Sigma$ and $h(x)$ is its determinant. Therefore, geometrical quantities as the volume of a region or the area of a surface can be written in terms of the electric field. In Loop Quantum Gravity these quantities correspond to operators having discrete spectra and eigenstates given by spin network states \cite{Rovelli:1994ge,Bianchi:2008es}. The area and the volume operator provide an interpretation of spin network states in terms of quantum geometries: quanta of volume are associated to regions dual to nodes of the spin network graph, and quanta of area are associated to surfaces dual to links of the graph\footnote{In \cite{Bianchi:2008es} this picture is further developed in order to take into account the length operator, too.}. Within the setting described in this paper, a cellular decomposition is present from the very beginning and a spin network with graph dual to it arises as the appropriate tool to describe gauge- and diffeomorphism-invariant functionals of the connection. It would be interesting to understand if, in this setting, operators build out of the flux of the electric field can attach a geometric meaning to the cells of the decomposition. This perspective was first proposed in \cite{Barbieri:1997ks} and is currently under study.
\end{itemize}
We point out some key differences between the spin network states discussed in this paper and the ones used in ordinary Loop Quantum Gravity. In Loop Quantum Gravity, embedded spin networks are not diffeomorphism invariant. In fact, states with diffeomorphic graphs which do not coincide are orthogonal. However, once the diffeomorphism constraint is imposed, most of these states are identified: abstract spin networks (also called s-knots) actually depend only on the diffeomorphism equivalence class of the graph and on the $SU(2)$ labels attached to it. The diffeomorphism equivalence class of graphs knows about the connectivity of the graph, about its knotting with the non-trivial topology of the manifold, about the self-knotting of the graph, and about some continuous parameters (moduli) related to the tangents to the links at each node of the graph \cite{Grot:1996kj}. These continuous parameters are due to the assumed differentiable structure at the nodes and disappear if a larger class of diffeomorphisms (the so called $\text{Diff}^*$) is considered \cite{Fairbairn:2004qe,Rovelli:2004tv}. On the other hand, the spin networks considered in our analysis arise as a tool for describing functions on the moduli space of locally-flat connections. As a result, even if the graph of our spin networks is embedded, the state they define is actually diffeomorphism-invariant. Moreover, they do not depend on the self-knotting of their graph, nor on the continuous moduli associated to the differentiable structure at the nodes. The only thing they actually depend on is their knotting with the manifold $\Sigma'$, i.e. with the network of defects and with the topology of the original manifold $\Sigma$. From this point of view, they are closer to the use that is made of spin networks as boundary states within the context of spin foams. This fact leads us to the next two points.
\begin{itemize}
\item \emph{Dynamics and spin foams}. The construction discussed is purely kinematical. A natural tool for implementing its dynamics is provided by spin foams. Spin foam models \cite{Baez:1997zt} can in fact be considered as topological field theories of the $BF$ type \cite{Horowitz:1989ng,Birmingham:1991ty,Cattaneo:1995tw,Baez:1999sr} defined on a manifold with defects. This statement is supported by the following two observations: (i) at the classical level, General Relativity can be formulated as a topological $BF$ theory with a set of constraints on the $B$ field \cite{Plebanski:1977zz}; at the quantum level, in spin foams, these constraints are imposed \emph{only} on the $2$-boundary of the $4$-cells of a simplicial complex. Thus a $2$-skeleton of defects where the constraints are imposed is identified. A second remark is that (ii) the currently studied spin foam models \cite{Barrett:1997gw} are based on the quantization of a version of Regge simplicial gravity \cite{Regge:1961px}. At the classical level, Regge action can be thought as the evaluation on a locally-flat metric of the Einstein-Hilbert action for a manifold with defects. The boundary of the manifold is given by the defect-network and Regge action consists in fact solely of the Gibbons-Hawking-York boundary term \cite{Gibbons:1976ue}, i.e. of the integral of the area density times the extrinsic curvature at the defect. The theory has finitely many degrees of freedom which are in fact associated to the defect-network. 
\item \emph{Infinitely many degrees of freedom}. Spin foam models provide a sum over histories. An history can be obtained from a cellular decomposition of a $4$-manifold $\mc{M}$. Then we can introduce a new manifold $\mc{M}'$ obtained subtracting from the original manifold the $2$-skeleton of the cellular decomposition. As a result, a foliation of $\mc{M}'$ results in a family of $3$-manifolds $\Sigma'$ with different networks of defects. A model of this kind allows transitions between Hilbert spaces $\mc{K}'$ associated to different defect-networks. This fact points to the need of considering a larger (but still separable) Hilbert space $\mc{F}$ obtained \`a la Fock \cite{Streater:1989vi}. The analogous of the $n$-particle state space would be the Hilbert space $\mc{K}_n$ associated to defect-networks having moduli space of dimension $n$, so that $\mc{F}$ is the closure in norm of the span of states belonging to $\mc{K}_0\oplus\mc{K}_1\oplus\mc{K}_2\oplus\cdots$. The Hilbert space $\mc{F}$ is the one proper to a system with infinitely many degrees of freedom\footnote{We point out a difference with respect to ordinary Loop Quantum Gravity \cite{Ashtekar:2004eh} where the infinitely many degrees of freedom are associated to a `quantum' configuration space consisting of distributional polymeric connections. On the other hand -- in the scenario we are describing -- the infinitely many degrees of freedom are associated to quantum configurations consisting of connections having \emph{distributional polymeric magnetic field}.}. 
\end{itemize}

The idea that purely topological theories might play an important role at the Planck scale has already been noticed in the past\footnote{See for instance the last sections of \cite{Witten:1988ze} and of \cite{Horowitz:1989ng}.}. In the first pioneering papers on Loop Quantum Gravity \cite{Rovelli:1987df}, topological invariants played a key role (see also \cite{DiBartolo:1999ee}). More recently, 't Hooft \cite{'tHooft:2008kk} introduced a locally finite model for 4d gravity which is topological and has dynamical string defects. It would be interesting to explore its quantum kinematics within the setting described in this paper\footnote{A different approach to the quantization of BF theory with stringy defects has been proposed in \cite{Baez:2006sa}.}. 

We conclude this section with a remark about an intriguing long-sighted perspective:
\begin{itemize} 
 \item \emph{Locality and effective description}. Topological field theories have only global degrees of freedom. However, in presence of defects, some of the degrees of freedom are associated to the defect-network. As a result, there is a finite number of degrees of freedom associated to each region of the manifold, thus leading to a notion of locality. The discrete spectrum of quantum-geometry operators sets the scale of the region. This fact points to an appealing scenario for Quantum Gravity where the theory has no trans-planckian degrees of freedom because it is topological (and therefore finite) at small scales, while at larger scales it has finitely many degrees of freedom which can be described effectively in terms of a local quantum field theory. Loop Quantum Gravity may provide a realization of this scenario.
\end{itemize}

\section{Conclusions}
In this paper we have studied the state space of a theory of locally-flat connections on a manifold with defects. We briefly summarize our analysis: 
\begin{itemize}
	\item[(i)] The space $\mc{A}_f/\mc{G}$ of locally-flat connections modulo gauge transformations is finite-dimensional. Therefore a state, i.e. a gauge- and diffeomorphism-invariant functional of the connection, is actually given by an ordinary function of a finite number of parameters (moduli). A scalar product between states is introduced in terms of a functional integral over $\mc{A}_f/\mc{G}$. The functional integral reduces to an ordinary integral over the space of moduli with a non-trivial measure. The final result is given in equation (\ref{eq:measureFP}). The measure consists in the product of the Jacobian from gauge-fixed connections to moduli, times a Faddeev-Popov determinant expressed in terms of the moduli.
	\item[(ii)] The field theoretical derivation of the scalar product can be compared to an a priori unrelated construction which uses the techniques proper to Loop Quantum Gravity. This construction consists in considering states labeled by an embedded graph and build out of the holonomy of the connection. As the connection is locally-flat, these states actually depend only on the knotting of the graph with the network of defects. We consider a class of graphs $\Gamma'$ having edge-path group $\pi(\Gamma')$ isomorphic to the first homotopy group $\pi_1(\Sigma')$ of the manifold $\Sigma'$. In the particular case of a defect-network arising as the $1$-skeleton of a cellular decomposition, the graphs $\Gamma'$ is dual to the decomposition. These states correspond to specific functions on moduli space. Moreover they suggest a natural proposal for the choice of measure involved in the definition of the scalar product: this is the Haar measures for group elements associated to the links of the graph. Spin network states with graph $\Gamma'$ provide an orthonormal basis of the Hilbert space. We notice that the group theoretical measure induces a non-trivial measure on moduli space via Weyl integration formula.
	\item[(iii)] We argue that the measure on moduli space obtained via the field theoretical construction and the one obtained from the product of Haar measures coincide. Proving this conjecture would establish spin network states as an orthonormal basis of the Hilbert space $\mc{K}'$ build via the field theoretical construction.
	\item[(iv)] We provide a test of this conjecture in the case of a single defect-line. The moduli are identified with the flux of the magnetic field through the defect. The Faddeev-Popov determinant is computed explicitly and the result is shown to coincide with the measure on the class-angle induced by the Haar measure. The physical picture proper to the Aharonov-Bohm effect arises.
\end{itemize} 
The interest in this analysis lies in its relation with Loop Quantum Gravity and in the new perspectives it opens.

\section*{Acknowledgments}
\hspace{1.5em}Thanks to Giorgio Immirzi, Pietro Menotti and Carlo Rovelli for many insightful discussions and for encouragement. 
I wish also to thank Benjamin Bahr, Marc Knecht, Antonino Marcian\`o, Leonardo Modesto, Roberto Pereira, Daniele Pranzetti, Alejandro Perez, Michael Reisenberger and Matteo Smerlak for comments and suggestions on a preliminary version of this work.

\appendix

\section{First homotopy group of the manifold $\Sigma'$}\label{app:pi1}
We recall the definition of the first homotopy group and discuss its presentation \cite{Singer:1976tg}.  Let $\Sigma'$ be a connected $3$-manifold. A loop based at $x_0$ in $\Sigma'$ is a closed path starting and ending at the base point $x_0\in\Sigma'$. Two loops $\gamma_1$ and $\gamma_2$ based at $x_0$ are said to be \emph{homotopic} if one can be continuously deformed into the other without moving the base point. Homotopy is an equivalence relation: we denote $[\gamma]$ the class of loops homotopic to $\gamma$. The set of all homotopy classes of loops based at $x_0$ is denoted $\pi_1(\Sigma',x_0)$. This set is indeed a group, the fundamental group or first homotopy group, with the product given by composition of loops, the inverse given by the loop traversed in the opposite direction, and the identity given by the class of contractible loops. The first homotopy group depends both on the topology of the manifold $\Sigma'$ and on the base point $x_0$. As we have assumed that the manifold $\Sigma'$ is connected, the dependence on the base point can be dropped as different base points correspond to isomorphic groups. 

In this paper, the manifold $\Sigma'$ arises from a manifold $\Sigma$ and a network of defects $\mc{D}$, so that $\Sigma'=\Sigma-\mc{D}$. We point out that the group $\pi_1(\Sigma-\mc{D})$ is well-studied for some specific choices of $\mc{D}$. When the defect-network $\mc{D}$ consists of a collection of non-intersecting loops, the group $\pi_1(\Sigma-\mc{D})$ is called the \emph{knot} group. A convenient presentation (a set of generators and defining relations) for the knot group is provided by Wirtinger presentation \cite{Burde:2003kn}. 
In the paper we are mainly interested in the case of a network $\mc{D}$ arising as the $1$-skeleton of a cellular decomposition of the manifold $\Sigma$ \cite{Fritsch:1990ct}, $\mc{D}\equiv \mc{C}_1(\Sigma)$. In this case the first homotopy group of the manifold $\Sigma'=\Sigma-\mc{D}$ is discussed by Regge in \cite{Regge:1961px} (see also \cite{Frohlich:1994gc}). We have a generator for each defect-line and a number of relations for each point where defect-lines join. A related presentation can be obtained via the following construction. We build a graph $\Gamma$ dual to the cellular complex, i.e. topologically equivalent to the $1$-skeleton of the dual to $\mc{C}(\Sigma)$. Then we consider the edge-path group $\pi(\Gamma)$ of the graph $\Gamma$. Its elements are closed paths on the graph $\Gamma$, based at a vertex of the graph. The edge-path group of $\Gamma$ is isomorphic to the first homotopy group of $\Sigma-\mc{C}_1(\Sigma)$, $\, \pi(\Gamma)\sim \pi_1(\Sigma-\mc{C}_1(\Sigma))$, as can be shown considering a fattening of the graph $\Gamma$. Then, the standard presentation of $\pi(\Gamma)$ can be used \cite{Singer:1976tg}.

\section{The Jacobian from gauge-fixed connections to moduli}\label{app:J}
In order to compute the Jacobian from the gauge-fixed connections to the moduli we can use  a procedure similar to the one introduced by Polyakov in \cite{Polyakov:1981rd}. We consider the space of gauge-fixed locally-flat connections. A point in such space is a flat connection $\bar{A}^{m_r,g_0}$ satisfying a gauge-fixing condition $\chi^i=0$ and labeled by the moduli $m_r$ and by a global $SU(2)$ transformation $g_0$ which is left unfixed by the condition  $\chi^i=0$. The tangent space at a point $\bar{A}^{m_r,g_0}$ is spanned by vectors $\delta\bar{A}^{m_r,g_0}$. Such vectors can by decomposed in terms of the variation of the moduli $\delta m_r$ and the variation of the global $SU(2)$ element $\xi^i=(\delta g_0)^i$,
\begin{equation}
\delta \bar{A}_a^i=\frac{\partial \bar{A}_a^i}{\partial m_r}\delta m_r+{\eps^i}_{jk} \bar{A}_a^j\,\xi^k\;.
\end{equation} 
We can introduce a scalar product on this tangent space:
\begin{equation}
(\delta \bar{A},\delta \bar{A})=\int d^3x \,\delta^{ab}\delta_{ij}\;\delta \bar{A}_a^i\; \delta\bar{A}_b^j\;.
\end{equation}
Such scalar product involves the auxiliary metric $\delta_{ab}$ introduced via the gauge-fixing condition. The scalar product can be used to provide a Gaussian normalization of the formal measure on gauge-fixed connections and define the Jacobian $J(m_r)$,
\begin{equation}
1=\int d(\delta \bar{A}) \;\exp -(\delta \bar{A},\delta \bar{A})\;
=\;J(m_r)\;\int d\,\delta m_r \int d\xi\;\exp -(\delta \bar{A},\delta \bar{A})\;.
\label{eq:def J}
\end{equation}
The Jacobian is independent of the group element $g_0$ as the scalar product is invariant under global $SU(2)$ transformations.

In order to clarify the construction we compute the Jacobian $J(\phi)$ in the simple case of a single line defect of section \ref{sec:line}. We write the magnetic flux $\Phi^i$ in terms of its modulus $\phi$ and a given direction $\bar{v}^i$, $\,\Phi^i=\phi\;\bar{v}^i$, so that
\begin{equation}
\bar{A}^{\phi,g_0}_a=\frac{\phi}{2\pi}\alpha_a\, g_0^{-1} (\bar{v}^i\tau_i) g_0\;.
\end{equation}
The tangent vector at a configuration is simply given by
\begin{equation}
\delta \bar{A}_a^i=\frac{\alpha_a}{2\pi}\big(\bar{v}^i\,\delta \phi+\phi {\eps^i}_{jk}\bar{v}^j\xi^k\big)\;.
\end{equation}
As a result we have that, defining $c$ the integral of $(2\pi)^{-2}\delta^{ab}\alpha_a \alpha_b$, the scalar product $(\delta \bar{A},\delta \bar{A})$ is given by
\begin{equation}
(\delta \bar{A},\delta \bar{A})=c\;\big((\delta \phi)^2+\phi^2 \,\xi^k P_{kl}^{\bot} \xi^l\big)
\end{equation}
with $P_{kl}^{\bot}$ the projector in $\mbb{R}^3$ orthogonal to the vector $\bar{v}^i$. Using relation (\ref{eq:def J}) we find
\begin{equation}
J(\phi)=\sqrt{{\det}_\bot (\phi^2 P_{kl}^{\bot})}=\phi^2
\end{equation}
which matches with the simpler derivation discussed in section \ref{sec:line}.

\section{The Faddeev-Popov determinant in the case of a line defect}\label{app:FP}  
Here we compute the Faddeev-Popov determinant $\Delta_{\text{FP}}(\phi)$ used in section \ref{sec:line}. The gauge-fixing function we consider is the Coulomb-like gauge $\chi^i=\partial^a A_{a}^i=0$. The variation of $\chi^i$ with respect to a gauge transformation defines an operator $K$ which depends on the connection. The quantity $\Delta_{\text{FP}}$ is given by the functional determinant of this operator. Computing it is in general hard unless one resorts to perturbative techniques. In the case studied in this paper, however, the connection involved is locally-flat so that the moduli space is finite dimensional. Thus the Faddeev-Popov determinant is an ordinary function of few parameters -- the moduli -- and we can hope to determine it non-perturbatively, at least in simple cases. In particular, in the case of a single line defect discussed in section \ref{sec:line}, the moduli space is one-dimensional and the Faddeev-Popov determinant $\Delta_{\text{FP}}(\phi)$ is in fact an ordinary function on the interval $[0,2\pi]$. It is given by the (appropriately regularized) product of the eigenvalues of the differential operator $K(\phi)$. In the following we study this operator.

We work in cylindrical coordinates $(z,r,\theta)$. In order to simplify the analysis of the boundary conditions to be imposed on the defect line at $r=0$, we actually focus on a improved gauge-fixing function $\chi^i=-r^2 \partial^a A_{a}^i=0$. This improvement does not change the $\phi$-dependence of the Faddeev-Popov determinant. We have that the differential operator $K_{ij}(\phi)$ is given by
\begin{equation}
K_{ij}(\phi)=\left.\frac{\delta \chi^i}{\delta \xi^j}\right|_{\bar{A}^{\phi}}=-\delta_{ij} (r^2\partial_z^2+r^2\partial_r^2+r\partial_r+\partial_\theta^2) - \eps_{ijk}\bar{v}^k \,\frac{\phi}{2\pi} \partial_\theta
\label{eq:K}
\end{equation}
where $\bar{v}^i$ is a given direction that we have fixed in the previous section. To derive (\ref{eq:K}) we have used the fact that $\partial^a \alpha_a=0$ and $\alpha_a\partial^a=r^{-2}\partial_\theta$. The operator is defined on functions $\xi^i(x)$ with $x\in \Sigma'\equiv\mbb{R}^3-l$ and boundary conditions at $r=0$
\begin{equation}
\left.\partial_z \xi^i\right|_{r=0}=0\quad,\quad  \left.\partial_r \xi^i\right|_{r=0}=0\;.
\label{eq:boundary condit}
\end{equation}
Such boundary conditions on gauge transformations guarantee that the affine part of the connection is unchanged on the defect line. The eigenvalue problem we have to solve is 
\begin{equation}
{K^i}_{j}(\phi) \xi^j=\lambda \xi^i 
\end{equation}
 Eigenfunctions can be found via separation of variables. We consider an ansatz of the form $\xi^i(z,r,\theta)=A(z)B(r) C^i(\theta)$ and introduce two separating constants, $k_z$ and $c_n$. We have that
\begin{align}
&\partial_z^2 A+k_z^2 A=0 \;,\label{eq:kz}\\
&\partial_\theta^2 C^i+\frac{\phi}{2\pi}{\eps^i}_{jk} \bar{v}^j \partial_\theta C^k-c_n C^i=0\;, \label{eq:cn}\\
&r^2 \partial_r^2 B+ r \partial_r B-(k_z r^2+c_n+\lambda) =0
\;.\label{eq:Bessel}
\end{align}
The first equation is solved by $A(z)=a_1 \cos k_z z\,+a_2 \sin k_z z$. The second equation is solved by the two periodic functions
\begin{align}
C_{n,+}^i(\theta)=&\; (\cos n \theta\,,\,\sin n \theta\,,\, 0)\,,\\
C_{n,-}^i(\theta)=&\; (\sin n \theta\,,\,\cos n \theta\,,\, 0)\,,
\end{align}
where the third component\footnote{Periodicity of $C_{n,\pm}^i(\theta)$ for $\phi\neq 0$ imposes that its component in the direction $\bar{v}^i$ vanishes.} corresponds to the direction $\bar{v}^i$, $n$ in an integer and the separating constant $c_n$ is given by
\begin{equation}
c_n=-n^2\pm\frac{\phi}{2\pi}n\;.
\end{equation}
The general solution of equation (\ref{eq:Bessel}) is a linear combination of modified Bessel functions
\begin{equation}
B(r)=b_1\, I_{\sqrt{c_n+\lambda}}(k_z r)+ b_2\, K_{\sqrt{c_n+\lambda}}(k_z r)\;.
\end{equation} 
The boundary conditions (\ref{eq:boundary condit}) impose that $k_z=0$ and $\sqrt{c_n+\lambda}=0$. As a result we have that eigenfunctions are labeled by the integer $n$ and a sign. The eigenvalues are
\begin{equation}
\lambda_{n,\pm}=n^2\pm\frac{\phi}{2\pi}n\;.
\end{equation} 
In the following we will denote them simply by $\lambda_n$, with $n=\pm 1, \pm 2,\cdots$. The eigenvalues are twice degenerate, the corresponding eigenfunctions being $\xi_{n,\pm}^i= C^i_{n,\pm}(\theta)$. 

The determinant of the operator $K(\phi)$ can be computed as a product of its eigenvalues (with the appropriate degeneracy), defined via zeta-function regularization. The dependence of the Faddeev-Popov determinant on the modulus $\phi$ can be obtained considering the ratio
\begin{equation}
\Delta_{\text{FP}}(\phi)=c\,\frac{\text{Det} K(\phi)}{\text{Det} K(0)}=c\,\prod_{n=\pm1,\pm2,\cdots}\frac{\big(\lambda_n(\phi)\big)^2}{\big(\lambda_n(0)\big)^2}\;,
\end{equation}
where $c$ is an undetermined constant.


\providecommand{\href}[2]{#2}\begingroup\raggedright\endgroup

\end{document}